\documentclass[prl,twocolumn,showpacs,amsmath,amssymb,aps,10pt]{revtex4-1}

\usepackage[pdftex]{graphicx}
\usepackage{hyperref}
\hypersetup{
 pdfnewwindow=true, colorlinks=true,
 linkcolor=blue, anchorcolor=blue,
 citecolor=blue, filecolor=blue,
 menucolor=blue, urlcolor=blue}
\usepackage{breakurl}

\begin{document}

\title{Surface Atom Motion to Move Iron Nanocrystals through
  Constrictions in Carbon Nanotubes under the Action of an Electric
  Current}

\author{Sinisa Coh}
\email{sinisa@civet.berkeley.edu} 
\author{Will Gannett}
\author{A. Zettl}
\author{Marvin L. Cohen} 
\author{Steven G. Louie} 
\affiliation{Department of Physics, University of California at
  Berkeley}
\affiliation{Materials Sciences Division, Lawrence Berkeley National
  Laboratory, Berkeley, California 94720, USA}

\date{\today}


\begin{abstract}
  Under the application of electrical currents, metal nanocrystals
  inside carbon nanotubes can be bodily transported. We examine
  experimentally and theoretically how an iron nanocrystal can pass
  through a constriction in the carbon nanotube with a smaller
  cross-sectional area than the nanocrystal itself. Remarkably,
  through in situ transmission electron imaging and diffraction, we
  find that, while passing through a constriction, the nanocrystal
  remains largely solid and crystalline and the carbon nanotube is
  unaffected. We account for this behavior by a pattern of iron atom
  motion and rearrangement on the surface of the nanocrystal. The
  nanocrystal motion can be described with a model whose parameters
  are nearly independent of the nanocrystal length, area, temperature,
  and electromigration force magnitude. We predict that metal
  nanocrystals can move through complex geometries and constrictions,
  with implications for both nanomechanics and tunable synthesis of
  metal nanoparticles.
\end{abstract}

\maketitle

Electric current induced movement of metal nanocrystals in and on
multiwall carbon nanotubes has been observed for many metals including
iron \cite{ref01,ref02,ref03,ref04}, copper \cite{ref05},
tungsten \cite{ref06}, indium \cite{ref07}, and gallium \cite{ref08}.
The direction of movement is directly related to the direction of the
applied current, and is often entirely reversible, i.e. the
nanocrystal can be moved back and forth by simply switching the
applied current polarity.  The speed of the nanocrystal within the
nanotube is dependent upon the applied current magnitude.  From an
applications view the mechanism is of great interest because it
provides an especially convenient method of controlling the nanocrystal's
position and motion with a single external electrical control
parameter.  Controlled movement of metal nanocrystals inside carbon
nanotubes could potentially be used for nanomachine
actuators \cite{ref09}, memory elements \cite{ref03}, or dispensing
small quantities of metals \cite{ref01} to a selected location.

The transport of metal nanocrystals within nanotubes is conventionally
demonstrated for nanotubes which have a relatively smooth, uniform
diameter hollow core, within which the nanocrystal can easily slide.
A critical question, however, is what happens when the nanotube core
contains a constriction smaller than the incoming nanocrystal
cross section.  The naive answer is that if the nanocrystal remains
solid it will be completely blocked by the constriction, while if it
is heated beyond its melting point and becomes liquid it might
(assuming surface tension energies can be overcome) squeeze through.
We here demonstrate how a metal nanocrystal, while remaining solid and
crystalline, can in fact be made to slip through a very small
constriction through which it should not geometrically fit.  The
squeezing mechanism is decidedly not one of severe deformation and
plastic flow, but rather a form of atomic level deconstruction at the
crystal’s trailing edge and reconstruction at the leading edge.
Indeed, this deconstruction and reconstruction of surface atoms is a
continual process even without a constriction:  It can be the dominant
mechanism by which the electrical current transports the metal nanocrystal
through any nanotube bore, smooth or not.

To demonstrate transport of iron nanocrystals through a carbon
nanotube constriction experimentally, we fabricate a two-terminal
nanotube device suitable for insertion into a high resolution
transmission electron microscope (JEOL 2010) with nanodiffraction and
dark field analysis capability.  Multiwall carbon nanotubes containing
iron nanoparticles are grown by pyrolysis of ferrocene in an inert gas
atmosphere at 1000$^{\circ}$C.  Such nanotubes often have naturally occurring
constrictions within their interior (e.g. where the number of tube
walls abruptly changes).  The nanotubes that contain iron nanocrystals
are then deposited onto thin silicon nitride membranes and electrical
contacts are formed using electron beam lithography.  The resulting
two-terminal nanotube device is driven with a dc electrical current
during TEM imaging, which allows observation and control of
nanoparticle motion in real time.  The nitride membrane platform
allows the same device to be measured multiple times and affords
mechanical stability during TEM imaging.

\begin{figure}[]
\centering\includegraphics{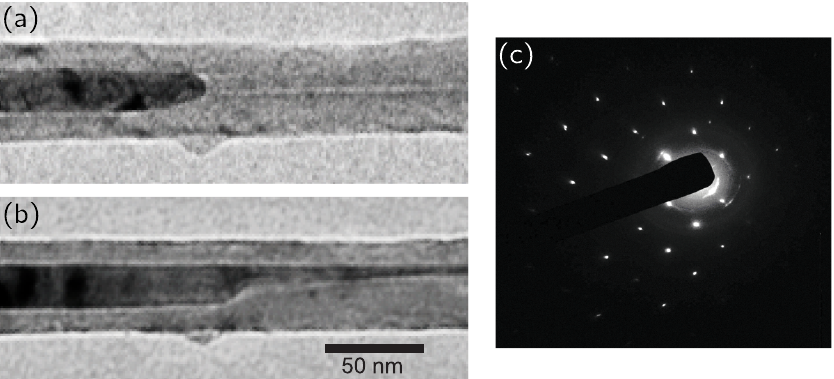}
\caption{Microscopy and diffraction of nanotube system. Panel (a)
  shows a transmission electron micrograph of an iron nanocrystal
  (darker contrast, spanning left half of image) inside a multiwalled
  carbon nanotube.  Panel (b) shows the same section of nanotube after
  a current has been applied, causing the iron nanoparticle to squeeze
  into the adjacent constriction.  The iron nanocrystal (darker
  contrast) now spans the full image width.  Panel (c) shows a
  diffraction pattern for a different iron nanoparticle while moving
  through a nanotube constriction, confirming its crystallinity.}
\label{fig:1}
\end{figure}

We observe that injecting the electrical current axially to the nanotube
causes the iron nanoparticle to move in the direction of the electron
flow.  The velocity varies with the current nonlinearly and the motion
of the nanoparticle is reversible, consistent with previous
observations \cite{ref03,ref04}.  We also observe the movement of iron
nanoparticles into and through narrow constrictions within the
nanotube, as demonstrated in Fig.~\ref{fig:1}.  Figure~\ref{fig:1}~(a)
shows a TEM image of a multiwall carbon nanotube with an approximately 45~nm
outer diameter.  The nanotube core on the left side of the image is
approximately 20~nm in diameter, and it is filled with iron (dark
contrast).  Midway along the axis of the nanotube there is clear
constriction, and the core reduces to about 5~nm.  In
Fig.~\ref{fig:1}~(a) the iron nanoparticle borders this constriction,
and the reduced inner diameter core is empty beyond.  Fig. 1b shows
the same nanotube region several seconds later; the iron nanoparticle
has advanced to the right and has infiltrated the narrow core region
beyond the constriction (also now dark contrast).  Numerous similar
iron infiltrations into and past constrictions are observed for
different samples.  In general, the current-driven iron nanoparticle
squeezes into the constriction and then continues to advance. If there is
sufficient space beyond the constriction, the entire iron nanoparticle
moves through the constrictions and emerges out the other side, and
continues to transport along the core of the nanotube.  In order to
determine the state (solid or liquid) of the iron during transport,
including while infiltrating the constriction, nanodiffraction
experiments are performed in situ.  Figure~\ref{fig:1}~(c) shows an
example (for a different iron nanoparticle) within a nanotube
constriction; the diffraction pattern is consistent with solid iron in
the bcc phase, even for the portion squeezing through the
constriction.  Additionally, real-time dark field imaging is performed
using one of the bcc iron diffraction spots, which confirms the
crystallinity of the iron during transport and squeezing though the
constriction.  The iron nanoparticle is solid, crystalline, and
lattice undeformed as it squeezes through the constriction.

Although the experiment is performed at room temperature, there is a
possibility that the nanotube with the iron nanoparticle is heated due to
Joule heating from the electrical current.  Using the parameters of
the experiment a detailed analysis \cite{ref10} reveals that the iron
nanocrystal of Fig.~\ref{fig:1}~(a,b) is at most at temperature
$T\approx$~440~K while it squeezes though the constriction.  This
temperature is well below the melting point of iron of this size
($T_{\rm melt}\approx 1800$~K), consistent with the nanodiffraction and
dark field imaging results always indicating a solid crystalline
state.

We now seek to understand why an iron nanocrystal can move through a
constriction in a carbon nanotube with a smaller cross-sectional area
than the nanocrystal itself.  We first examine the microscopic origin
of iron atom movement inside the carbon nanotube with a smooth bore and
then adapt the model to constrictions.  We perform a series of first
principles density functional theory \cite{ref11,ref12} calculations, followed
by a kinetic Monte Carlo simulation \cite{ref13,ref14}.

We use density functional theory calculations to compute the energy
barriers for iron atom diffusion in various environments. We examine
bulk diffusion, diffusion on different iron surface orientations, and
diffusion at the iron-carbon interface.  We also use density
functional theory calculations to obtain estimates of the iron-iron
and the iron-carbon binding energies.  Once we obtain these
parameters, we extract trends of diffusion energy barriers and binding
energies, and construct an algorithm to assign a diffusion barrier
height to arbitrary diffusion processes in iron.  Next, using this
algorithm for the assignment of diffusion barrier heights, we perform
kinetic Monte Carlo simulations of an iron nanocrystal inside a carbon
nanotube. The kinetic Monte Carlo method, unlike the static density
functional theory method, allows us to perform a time-evolution study
of the movement of the iron nanocrystal.

\begin{figure}[]
\centering\includegraphics{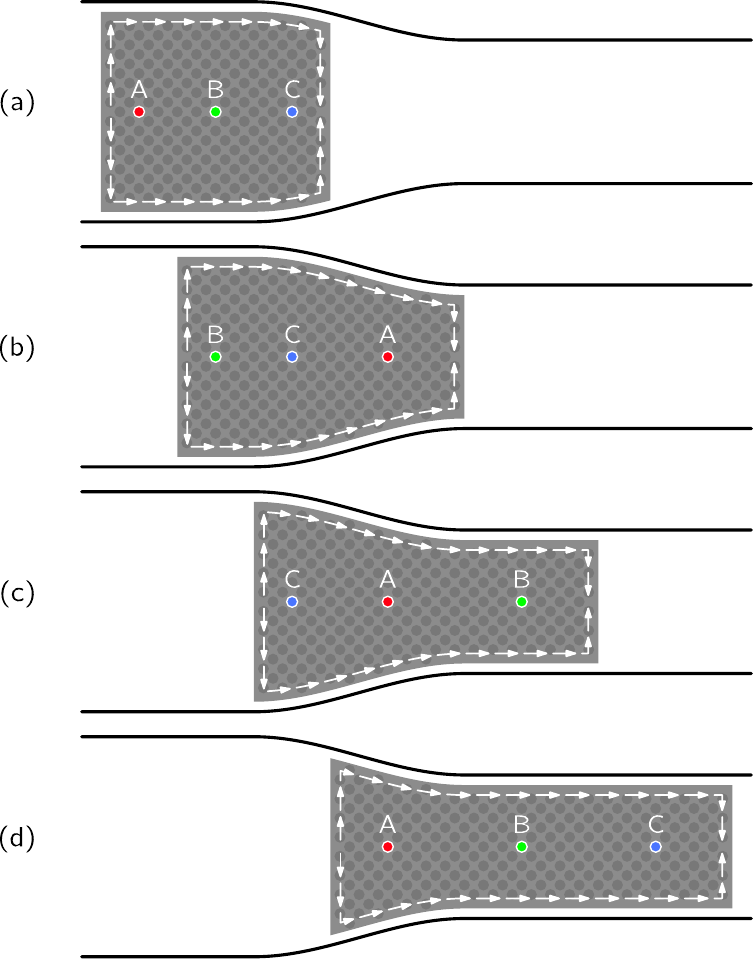}
\caption{(Color online.) Sketch of iron nanocrystal movement. Panels
  (a) through (d) schematically show four consecutive time snapshots
  of a solid iron nanocrystal (gray) moving through a constriction in
  a carbon nanotube (black). Atoms in the bulk of the nanocrystal
  remain stationary as long as they are in the bulk. Once iron atoms
  from the bulk are exposed to the end surface on the left side of the
  crystal, they quickly move along the nanocrystal surface and contact
  region with the carbon nanotube to the right end of the crystal
  (white arrows). For illustration purposes, instantaneous positions
  of three selected iron atoms are indicated with red, green, and blue
  color, and with capital letters A, B, and C. In each of the four
  panels the three selected iron atoms are in the bulk of the
  nanocrystal, illustrating the fact that compared to time spent in
  the bulk, surface movement is nearly instantaneous. Movement of iron
  atoms in the contact region with the carbon nanotube originates from
  the electromigration force. Additionally, this force creates a
  concentration gradient that drives the diffusion from the left
  (right) end of the crystal toward (away) the contact region (see
  also Fig.~\ref{fig:3}).}
\label{fig:2}
\end{figure}

\begin{figure}[]
\centering\includegraphics{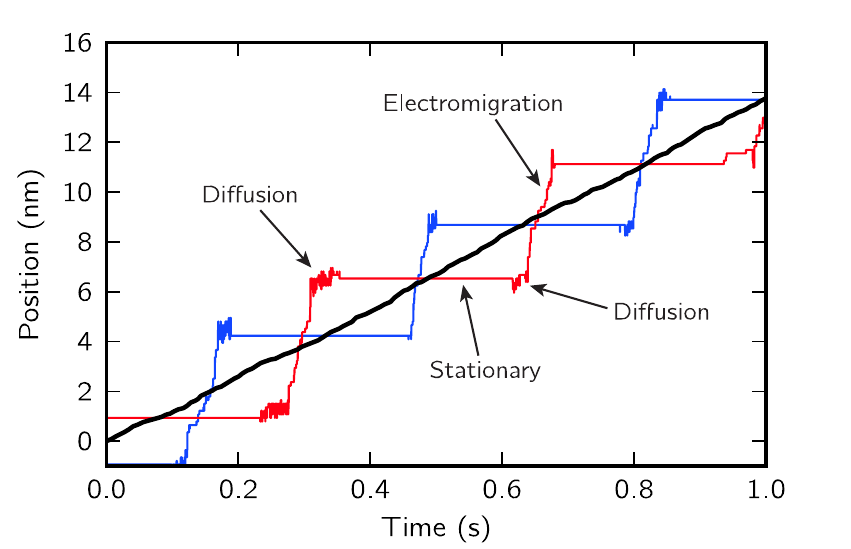}
\caption{(Color online.) Movement of individual iron atoms. Thin red
  and blue lines show the simulated positions along the carbon
  nanotube axis of two randomly selected iron atoms as a function of
  time. Thick black line shows the average position (center of mass)
  of all iron atoms in the simulation. The average position of the
  iron atoms is continuously increasing. On the other hand, individual
  iron atoms remain stationary most of the time. It is only when they
  are exposed to the surface that they move quickly across the entire
  length of the crystal by a combination of diffusion and
  electromigration forces (see also Fig.~\ref{fig:2}). This kinetic
  Monte Carlo simulation is performed at temperature of 600~K, the
  electromigration force on the iron atoms is 0.33~eV/nm, and the iron
  nanocrystal radius is $r_{\textrm{cyl}}=1.05$~nm while its length is
  $l=4.31$~nm.}
\label{fig:3}
\end{figure}

In the simulation, we assume that the electromigration force $F$
experienced by individual iron atoms is linearly proportional to the
current density $j$; $F=j K$ and we obtain a parameter $K$ by fitting
the results of our simulation to the experiment. (The linear
dependence of force $F$ on current density $j$ is consistent with an
electron wind force mechanism \cite{ref15,ref17}.)  Furthermore, we
assume that the electromigration force $F$ affects the iron atom
movement by increasing or decreasing the iron diffusion barrier height
according to the work done by $F$ along the atom diffusion path. The
depth of the contact region in which iron atoms are experiencing the
electromigration force $F$ does not affect the resulting speed of the
nanocrystal but only its instability toward breaking (a deeper contact
region produces an earlier onset of instability). Details of this
combined density functional theory and kinetic Monte Carlo calculation
will be presented elsewhere\cite{longPeg}. The microscopic results of
the kinetic Monte Carlo simulations are shown in Figs.~\ref{fig:2} and
\ref{fig:3}.  Figure~\ref{fig:2} schematically shows the nature of the
movements of the iron atoms. We find that for most of the time, any
given iron atom is stationary, as shown in Fig.~\ref{fig:3}. Once the
stationary bulk iron atoms are exposed to the surfaces they quickly
move from the left end of the nanocrystal (as in Fig.~\ref{fig:2}),
along the contact region with the carbon nanotube, toward the right
end of the crystal.  Since iron atoms are depleted from the left
surface, they expose a new layer of bulk atoms to the surface, which
then start to move in the same way.  Analogously, when these atoms
arrive to the right surface, they soon become buried under layers of
new incoming iron atoms, and thus they once again become part of the
bulk and become stationary.  The nanocrystal is deconstructed at the
left and reconstructed at the right, and the surface atoms overtake
their bulk counterparts.  Hence the nanocrystal moves, even though
virtually all of the atoms comprising the nanocrystal (i.e., all but
the surface atoms) remain stationary in the laboratory frame. A
somewhat related mechanism, but one involving the heating of an iron
nanocrystal and its chemical interaction with the carbon nanotube, was
recently proposed in Ref.~\onlinecite{ref04}.

The origin of the surface movement of iron atoms is twofold.  In the
contact region with the carbon nanotube, iron atoms move due to the
influence of the electromigration force from the applied current. On
the exposed iron surfaces on both the left and right sides of the
nanocrystal, movement is driven by the diffusion forces.  Diffusion
forces away from the left side and toward the right side of the
nanocrystal are originating in the iron atom concentration gradient
created by the electromigration force in the contact region with the
carbon nanotube.  The pattern of iron atom movement presented above
explains why the iron nanocrystal can move through a constriction
while remaining solid. If the iron nanocrystal were moving as a whole,
it would have to deform in order to go through a constriction.  On the
other hand, the mechanism we consider does not require deformation of
the crystal.  Instead, once the iron nanocrystal reaches a
constriction, new atoms are transported toward the region within the
smaller cross-sectional area, and they assemble there to form new
layers of iron atoms that adjust their cross-sectional area to match
the constriction.
 
\begin{figure}[]
\centering\includegraphics{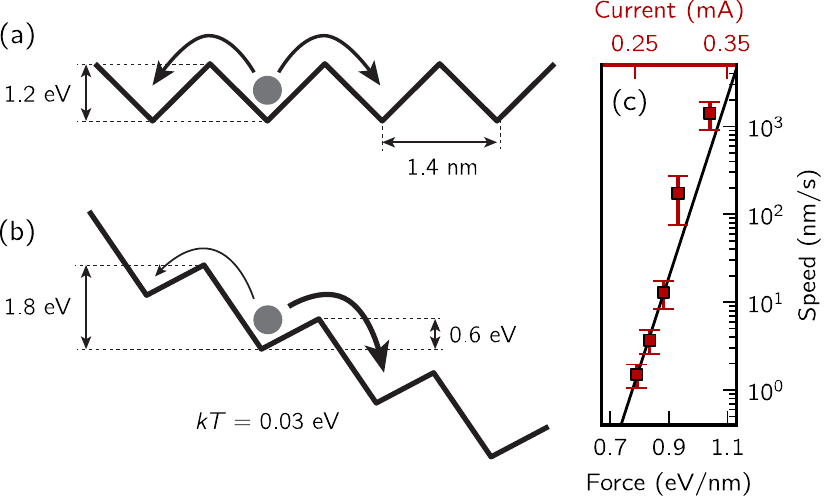}
\caption{(Color online.) The complex motion of an iron nanocrystal
  shown in Fig.~\ref{fig:2} and \ref{fig:3} can be simply modeled as
  the movement of a single particle in an effective periodic external
  potential with the barrier height $\tilde{B}$ of 1.2~eV and the
  period $\tilde{L}$ of 1.4~nm (assuming that the force experienced by
  single iron atom is applied to this effective particle). This is
  true regardless of the iron nanocrystal length, cross-sectional
  area, temperature, and magnitude of the electromigration force. When
  the electromigration force is not present (a), the particle behaves
  as if it is in an untilted washboard potential with equal energy
  barriers in left and right direction.  When the electromigration
  force $F$ is present (b) washboard potential is tilted with slope
  $F$ and the barrier heights become asymmetric, which prefers the
  motion of the particle along the direction of $F$. For typical
  experimental situation \cite{ref03}, asymmetry in the effective
  barriers heights is about half of untilted barrier height
  ($\pm0.6$~eV), while the temperature is much smaller than both, only
  0.03~eV. This regime of barrier heights explains the very strong
  observed exponential dependence of the iron nanocrystal speed on the
  electromigration force magnitude \cite{ref03,ref04}. Panel (c) shows
  iron nanocrystal speed on a logarithmic scale as a function of the
  electromigration force magnitude $F$ (black) and the net current
  through the carbon nanotube (red). Black line is a fit to
  Eq.~\ref{eq:1} with $T=350$~K, while red symbols are measurements
  from Ref.~\onlinecite{ref03}.}
\label{fig:4}
\end{figure}

Our theoretical modeling allows us to analyze the dependence of the
iron nanocrystal center of mass speed on various external parameters.
First, we find that the iron nanocrystal center of mass speed does
not depend on the nanocrystal length.  This observation is easily
explained by the fact that the electromigration force driven motion of
iron atoms in the contact region is much more effective than the
diffusion on the two ends of the nanocrystal.  Second, in our
simulations we find an exponential thermally activated dependence of
the center of mass speed $v$ on the electromigration force per iron atom
$F$ and the simulation temperature $T$
\begin{align}
  v = \tilde{v} \exp \left( - \frac{ \tilde{B}}{k T} \right) \sinh
  \left( \frac{\frac{1}{2} \tilde{L} F}{k T} \right).
  \label{eq:1}
\end{align}
Fitting this equation to the results of our simulation, we obtain
the following values of the $\tilde{v}$, $\tilde{B}$, and $\tilde{L}$
parameters:
\begin{align}                    
  \tilde{v} = 3.3 \textrm{~m/s,}
  \quad                          
  \tilde{B} = 1.2 \textrm{~eV,} 
  \quad                          
  \tilde{L} = 1.4 \textrm{~nm.} 
\label{eq:2}
\end{align}                      
The functional form given in Eq.~\ref{eq:1} is the same as that of a
single particle coupled to a thermal bath at temperature $T$ moving in
a periodic tilted washboard potential \cite{ref18} (shown in
Fig.~\ref{fig:4}(a,~b)) with period $\tilde{L}$, barrier height $\tilde{B}$,
and under the influence of a constant force $F$.  Finally, we find a
complex dependence of the center of mass speed on the cross-sectional
area of the iron nanocrystal (with an overall trend of decreasing
center of mass speed with increasing cross-sectional area). The origin
of this complex dependence comes from the fact that, depending on the
radius of the iron nanocrystal, one obtains different iron surface
morphologies with varying diffusion pathways in the contact region
with the carbon nanotube. Nevertheless, we find that varying the
cross-sectional area of the iron nanocrystal mostly influences the
value of the parameter $\tilde{v}$ in Eq.~\ref{eq:1}, while
$\tilde{B}$ and $\tilde{L}$ are essentially unchanged.  Since the
parameter $\tilde{v}$ appears only as a prefactor in Eq.~\ref{eq:1},
when fitting our model calculation to experiment \cite{ref03,ref04},
the precise value of parameter $\tilde{v}$ will be almost irrelevant
as compared to $\tilde{B}$ and $\tilde{L}$.  Thus, even though motion
of individual iron atoms in carbon nanotube is quite complex, the
effective speed of the entire iron nanocrystal can be simply modeled
over a wide range of external parameters as that of a single particle
in a tilted washboard potential.

Extrapolating our model calculation to the experimental regime of
parameters, we find good agreement with experiment \cite{ref03} using
a temperature of 350~K and a constant of proportionality
$K=0.18$~eV~nm/$\mu$A between the current density through the iron
nanocrystal and the electromigration force (see Fig.~\ref{fig:4}(c)).
To obtain this value of parameter $K$ ($0.18$~eV~nm/$\mu$A) we crudely
estimated the current density through the iron nanocrystal based on
the resistivity of {\it bulk} iron and graphite, the nanotube
geometry, and assuming a constant current density profile
perpendicular to the nanocrystal axis. There is thus a large
uncertainty in the assumed current density and hence $K$. We are
unaware of any previous theoretical or experimental estimates of
parameter $K$ in iron. Moreover theoretical estimates
\cite{ref16,ref17} of parameter $K$ for studied elemental metals vary
widely across the periodic table in magnitude and even in sign.
Furthermore, the value of parameter $K$ is very sensitive \cite{ref17} to
the atomic structure and differs for the self-electromigration and
electromigration of an impurity.  Interestingly enough, largest value
of parameter $K$ obtained in Ref.~\onlinecite{ref17} is that of iron
impurity electromigrating in aluminum (0.01~eV~nm/$\mu$A), which is
within an order of magnitude to the value we obtained.

Comparing values of $\tilde{B}$, $\frac{1}{2}\tilde{L}F$, and $k T$,
we find that, in a typical experimental situation \cite{ref03}, the
energy barrier $\tilde{B}$ is the largest, 1.2~eV.  The tilt of the
washboard potential due to electromigration force
($\frac{1}{2}\tilde{L}F$) equals $\pm0.6$~eV and is comparable to the
barrier $\tilde{B}$ itself, while the temperature energy scale $k T$
is an order of magnitude smaller than both, 0.03~eV.  This order of
energy scales (also shown on Fig.~\ref{fig:4}(a,~b)) together with
Eq.~\ref{eq:1} explains the origin of the experimentally found
extremely sharp onset of iron nanocrystal movement as a function of
applied electric current \cite{ref03,ref04} (see also
Fig.~\ref{fig:4}(c)).

External electric control of movement of iron nanocrystal is
interesting from both fundamental science and applications viewpoints.
The ability of an iron nanocrystal to remain crystalline while moving
through tubes and constrictions could allow for more stable operation
of nanoelectromechanical devices and opens up a possibility to explore
more complex geometries than the limited geometries discussed here.
Additionally, the intricate mechanism of iron nanocrystal movement
could be used to refine metallic nanoparticles.  For example, constant
regrowth of the iron nanoparticle during its movement in the carbon
nanotube could be used to remove contaminants and domain boundaries
or, potentially, to introduce them with very fine spatial control.
Additionally, it may be interesting to explore systems with diffusion
of multiple metallic species both theoretically and experimentally.

\begin{acknowledgments}
  We thank Gavi Begtrup for assistance with sample preparation and
  microscopy and David Strubbe for discussion. This work was supported
  by the Director, Office of Energy Research, Office of Basic Energy
  Sciences, Materials Sciences and Engineering Division, of the U.S.
  Department of Energy under Contract No. DE-AC02-05CH11231.
\end{acknowledgments}

\bibliography{pap}

\end{document}